\documentclass[prd,aps,twocolumn,nofootinbib]{revtex4-1}
\usepackage[latin1]{inputenc}
\usepackage[english]{babel}
\usepackage{graphicx}
\usepackage{color}
\usepackage{amsmath}
\usepackage{amssymb}
\usepackage{multirow}
\usepackage{caption}
\usepackage[parfill]{parskip}    
\usepackage{graphicx}
\usepackage{amssymb}
\usepackage{epstopdf}
\usepackage{color}
\usepackage{graphicx}
\usepackage{tipa}
\usepackage{hyperref}

\begin{document}

\title{\bf Newman-Penrose scalars and black hole equations of state}

\author{F. D. Villalba}
\email{fd.villalba10@uniandes.edu.co}
\affiliation{Departamento de F\'{\i}sica, Universidad de los Andes, CP 111711, Bogot\'a, Colombia}
\author{P. Bargue\~no}
\email{pedro.bargueno@ua.es}
\affiliation{Departamento de F\'isica Aplicada, Universidad de Alicante, Campus de San Vicente del Raspeig, E-03690 Alicante, Spain}
\author{A. F. Vargas}
\affiliation{School of Physics, University of Melbourne, Parkville VIC 3010, Australia}
\author{E. Contreras}
\affiliation{Departamento de F\'isica, Colegio de Ciencias e Ingenier\'ia, Universidad San Francisco de Quito, Quito, Ecuador.}

\begin{abstract}
In this work we explore the connections between Newman-Penrose scalars, including the Penrose-Rindler $\mathcal{K}$-curvature, with the equation of state of asymptotically Anti-de Sitter Reissner-Nordstr\"om black holes. After briefly reviewing the equation of state for these black holes from the point of view of both the Extended Phase Space and the Horizon Thermodynamics approaches, a geometric splitting is given for such an equation in terms of the non vanishing Newman-Penrose scalars which define the $\mathcal{K}$-curvature at the horizon. From this splitting, a possible thermodynamical interpretation is developed for such scalars in the context of the black hole thermodynamics approaches initially discussed. Afterwards, the square root of the Bel-Robinson tensor is employed to propose conditions at the horizons in terms of pressures or energy densities, which can be understood as alternative thermodynamical definitions of these surfaces.
\end{abstract}

\maketitle

\section{Introduction}
\label{IntroNP}

Among all consequences of Einstein's General Relativity (GR), black holes (BHs) populate the imagination of both physicists and mathematicians, with Hawking and Penrose theorems \cite{penrose1972techniques,hawking1973large} exemplifying this marriage in a superb way. In spite GR being a well-defined theory from the mathematical point of view, there are different approaches towards it, most of them based on techniques from differential geometry.
Among these approaches, the Newman-Penrose formalism (NP) for spin coefficients (SC) \cite{penrose1984spinors,penrose1984spinors2} enlightens GR, both from a fundamental and a practical/computational point of view, allowing interesting formal developments such as the Petrov classification and the Goldberg-Sachs theorem. One of the mathematical objects defined within the SC formalism, the Penrose and Rindler's complex curvature, $\mathcal{K}$ (see Ref. \cite{penrose1984spinors}, Eq. (4.14.20).), has been scarcely used along last decades, its main use coming from Hawking's topology theorem \cite{hawking1972black}, together with its generalizations \cite{woolgar1999bounded}, and from the definition of Hawking's quasilocal energy \cite{penrose1982quasi,hayward1994quasi,hayward2006gravitational}. In addition, Hayward rewrote \cite{hayward1994spin} the laws of BH dynamics in terms of the SC formalism, involving the $\mathcal{K}$-curvature in the description of the geometry of 2-surfaces and relating it to an energy density. Recently, in order to visualize space-time curvature via the so-called frame-drag vortexes, Thorne and coworkers have rewritten some features of the tidal and frame-drag fields in terms of $\mathcal{K}$ \cite{owen2011frame,nichols2011visualizing,zhang2012visualizing}.

In a different but related context, theoretical evidence has accumulated suggesting a deep relationship between gravitation, thermodynamics, and quantum theory \cite{jacobson1995thermodynamics,padmanabhan2010thermodynamical}. Recently, new perspectives \cite{kastor2009enthalpy} with respect to the thermodynamical role of the cosmological constant in the context of BH physics, first studied in \cite{teitelboim1985cosmological,henneaux1989cosmological}, led to the realization that BH thermodynamics is a much richer subject than previously thought. Namely, variations of this parameter for asymptotically anti-de Sitter (AdS) BHs, defined globally through approaches such as \cite{ashtekar1984asymptotically}, allow for the introduction of readily defined pressures, volumes, phase behaviors, etc., in such settings. The analysis of the resulting thermodynamics leads to the {\it extended phase space} (EPS) approach for BH thermodynamics (for further details, see \cite{kubizvnak2017black}). As a general feature, here we remark that asymptotically AdS BHs are found to be quite analogous to Van der Waals (VdW) fluids within this context.

Now well, to further pursue the aforementioned connections, in this work we provide a thermodynamic interpretation for the $\mathcal{K}$-curvature, together with other relevant scalars defined in the SC framework, in terms of the equation of state (EoS) of AdS-Reissner-Nordstr\"om (RN) BHs. This interpretation is motivated by a geometric splitting of such equation that let us associate certain combination of Newman-Penrose scalars to each term of the EoS. Interestingly, we find that the EoS corresponds intrinsically to an important theorem that related $\mathcal{K}$ with the Gaussian curvature of a 2-surface, applied to the BH horizon. The identification is performed within the EPS framework and then a comparison with another approach to BH thermodynamics, the so-called Horizon Thermodynamics (HT) \cite{padmanabhan2010thermodynamical}, is considered. Although it is clear that a thermodynamic interpretation for the NP scalars could be done directly from the laws of BH dynamics, which can be interpreted as thermodynamic under certain assumptions such as stationarity, it is not straightforward to assign the role of pressure to any quantity within this context \cite{dolan2011pressure}. Therefore, it is necessary to include prescriptions for the identification of pressure such as those present in EPS or HT, and this fact can be seen as a justification for our approach. Although much of our discussion is framed in the context od AdS-RN BH, mainly because they are the most general static solution for an Einstein-Maxwell system, there are some features of our results that can be extended to other static settings in a straightforward way, and this possibility is discussed along the text.

After presenting the identifications in EPS and HT, we venture to propose an additional identification for NP scalars in terms of pressures {\it uniquely}, taking into account the preceding results and a definition of pressure based on the {\it square root of the Bel-Robinson} (SQBR) tensor \cite{bonilla1997some}. To end, we show that the $\mathcal{K}$-curvature of the horizon can be decomposed in terms of the SQBR tensor and the aforementioned \textit{pressures} associated now with the Maxwell and AdS sector. An equivalent proposal is considered in terms of energy densities. We follow these findings with a discussion of their plausibility and their relation with previous research in the literature.

The manuscript is organized as follows. Section \ref{BHES} briefly summarizes both the Extended Phase Space (EPS) and Horizon Thermodynamics (HT)
approaches, from which the EoS for an AdS-RN BH is derived. The geometric splitting of the EoS in terms of the $\mathcal{K}$-curvature, which is one of our main findings, is discussed in Section \ref{GS}. while Section \ref{GSGEM} introduces a thermodynamical condition at (or defining) the BH horizons in terms of pressures associated with $\mathcal{K}$ by considering the SQBR tensor. Afterwards, Section \ref{Discussion} is devoted to discuss our results and possible future work, with some additional final remarks given in Section VI. We use units where $\hbar=c=k_B=G=1$, and our signature is $(+---)$. Conventions regarding Riemann tensor, Einstein equations, and the definition of scalars follow those of Penrose and Rindler \cite{penrose1984spinors}.

\section{Black Hole Equations of State}
\label{BHES}
To provide a self-contained context for our results, this section is devoted to review the main ideas leading to the concept of BH EoS using both the EPS and HT approaches. Emphasis will be given to the construction and interpretation of the EoS for the AdS--RN BH.
\subsection{Extended Phase Space}
\label{EPS}

In the EPS approach \cite{kastor2009enthalpy,dolan2011cosmological,dolan2011pressure,cvetivc2011black,dolan2012pdv,kubizvnak2015black,mann2016chemistry,dolan2015black}, also known as \textit{Black Hole Chemistry} \cite{kubizvnak2017black}, we consider a BH spacetime with an AdS background, where the cosmological constant, $\lambda$, is allowed to take different values. By comparing the asymptotically AdS spacetime with cosmological constant $\lambda$ with another spacetime of the same class, with corresponding constant $\lambda+d\lambda$, the first law can be written as $dM = TdS+V_{t}dP_{\lambda}+\phi dQ+...$, where $V_{t}$ is a \textit{thermodynamical volume} \cite{kubizvnak2017black} conjugate to the pressure $P_{\lambda}$, which is defined as $P_{\lambda} = -\frac{\lambda}{8\pi}$. An important consequence of this extended first law is that the mass $M$ is identified with the thermodynamical enthalpy $H$,
\textit{i.e: $H(P,S) = M$} \cite{kastor2009enthalpy}, instead of the internal energy $U$, as it can be read from the intensive and extensive variables
 of the new first law. Therefore, the corresponding conjugated variables are given by $T = \left(\frac{\partial M}{\partial S}\right)_{P_{\lambda}}$ and $V_{t} = \left(\frac{\partial M}{\partial P_{\lambda}}\right)_{S}$, respectively, as expected from the usual thermodynamical theory.

One of the main advantages of the EPS approach is the possibility of readily obtaining EoS for different AdS-BHs. For instance, the EoS for the AdS-Reissner--Nordstr\"om (AdS-RN) BH can be obtained as follows. The AdS-RN BH, in four dimensions $D=4$ and written in conventional coordinates, is characterized by the metric
\begin{equation}
ds^2=f(r)dt^2-f^{-1}(r)dr^2-r^2 d\Omega^2,
\end{equation}
where $d\Omega^2$ is the usual metric for the two--sphere and

\begin{equation}
f(r) = 1-\frac{2M}{r}+\frac{Q^{2}}{r^{2}}+\frac{r^{2}}{l^{2}},
\label{metricfunc}
\end{equation}

where the cosmological constant in terms of the AdS radius, $l$, is given by $\lambda = -\frac{3}{l^{2}}$.
The BH horizon, $r_{+}$, is defined from $g^{rr}(r_{+})=f(r_{+}) = 0$. From this horizon condition, the BH mass $M$ can be expressed in terms of the parameters $(Q,r_{+},l^{2})$, as

\begin{equation}
M = \frac{r_{+}}{2}\left(1+\frac{Q^{2}}{r^{2}_{+}}+\frac{r^{2}_{+}}{l^{2}} \right).
\label{Mass}
\end{equation}

As commented before, the importance of Eq. (\ref{Mass}) lies in the fact that it can be identified with the enthalpy, as argued within the EPS approach. Therefore, we can obtain both the temperature and the thermodynamical volume of the AdS-RN BH by the standard expressions $T = \left(\frac{\partial M}{\partial S}\right)_{P_{\lambda}}$ and $V_{t} = \left(\frac{\partial M}{\partial P_{\lambda}}\right)_{S}$ which explicitly yield

\begin{equation}
T = \frac{3r^{4}_{+}+l^{2}(r^{2}_{+}-Q^{2})}{4\pi r^{3}_{+} l^{2}}.
\label{TempRNAdS}
\end{equation}

and, after noting that $l^{-2} = 8 \pi P_{\lambda}/3$,

\begin{equation}
V_{t} = \frac{4 \pi r^{3}_{+}}{3},
\label{VolRNAdS}
\end{equation}
respectively.

Having already defined the pressure $P_{\lambda}$, and obtained the temperature $T$ and the thermodynamical volume $V_{t}$, an EoS can be constructed. In order to do this, let us define a specific volume $v=2 r_{+}$ ($l_{p}=1$). Then, Eq. (\ref{TempRNAdS}) can be written as a VdW EoS for the AdS--RN BH in terms of this specific volume as \cite{kubizvnak2017black},

\begin{equation}
P_{\lambda} = \frac{T}{v}-\frac{1}{2\pi v^{2}}+\frac{2Q^{2}}{\pi v^{4}}.
\label{VdWRN}
\end{equation}

Let us recall that the specific volume, in the context of usual thermodynamics, is given by $v = \frac{V_{t}}{\bar{N}}$, where $\bar{N}$ should be understood as a {\it number of particles}. In fact, we can associate a number of particles to the horizon from a thermodynamic point of view, following in spirit the ideas within \cite{vargas2018sads,padmanabhan2010equipartition,padmanabhan2010equipartition2,padmanabhan2010surface,padmanabhan2016momentum}, from which an interpretation of AdS-RN BHs in terms of a VdW gas can be proposed.

Provided we know the thermodynamical volume is given by Eq. (\ref{VolRNAdS}), then the particle number $\bar{N}$ should be proportional to $r^{2}_{+}$. This sets the proportionality constant as $4\pi /6$. Even more, when demanding consistency with $v = 2r_{+}$, then the number of particles $\bar{N}$ can be written as $A/6$, where $A$ is the area of the horizon, which could be interpreted as a realization of the Holographic Principle. Finally, with the help of $\bar{N}$ and the thermodynamical volume, $V_{t}$, the EoS for the AdS--RN BH formally coincides with that of a VdW EoS (including a second virial term), reading
\begin{equation}
\label{eoseps}
P_{\lambda}=\frac{\bar N T}{V_{t}}-\frac{1}{2\pi}\frac{\bar N^2}{V_{t}^2}+\frac{2 Q^2}{\pi}\frac{\bar N^{4}}{V^{4}_{t}}.
\end{equation}
From this equation, it follows that a corpuscular microscopical interaction model could be proposed to provide an statistical mechanical foundation to this result. This possibility has been recently addressed in Ref. \cite{vargas2018sads},
in which both the equation of state given by Eq. (\ref{eoseps}) and the Bekenstein-Hawking entropy of a $D$-dimensional AdS-RN BH are recovered, using techniques from statistical mechanics and employing certain heuristic
gravitational constraints.

\subsection{Horizon thermodynamics}
\label{HT}

The origin of HT was the realization \cite{padmanabhan2002classical} that the Einstein equations on the horizon
of spherically symmetric spacetimes can be interpreted in terms of the first law of thermodynamics. This relevant observation has been extended
to other cases corresponding to different gravitational theories and symmetries (for a review see \cite{padmanabhan2010thermodynamical,padmanabhan2013lanczos}),
and also to the study of the thermodynamics of null surfaces \cite{chakraborty2015thermodynamical}. The basic idea of HT is based on the identification
\begin{equation}
P_{tot}=T^{r}_{\,\,r},
\end{equation}
being $T_{\mu\nu}$ the energy-momentum tensor of the complete matter sector (including a possible cosmological
constant) evaluated at the horizon. Then, under the assumption of an Euclidean (thermodynamic) volume for the BH, the radial Einstein equation can be interpreted as an EoS, $P_{tot}=P_{tot}(T,V)$ for spherically symmetric AdS BHs \cite{hansen2017universality}. Interestingly,
this EoS does not depend on the specific form of $g_{tt}=g^{rr}$ and the specific matter content is relevant only when interpreting the results. Even more, the authors of Ref. \cite{hansen2017universality} derive the first law of HTs for Lovelock-Lanczos theories in the form $dE=TdS-P_{tot}dV$, where $E$ is an energy associated with the BH whose meaning is discussed below. In addition, the horizon enthalpy and Gibbs energy are defined according to the standard prescriptions $G=E-TS+PV$ and $H=G+TS$.

In the case of AdS--RN BHs we are interested in, the main differences between the EPS and HT approaches are:
(i) the work term $\phi dQ$ which is present in the EPS approach contributes, within HT, to the total pressure associated to the matter fields; (ii) in HT, the BH volume is assumed to be the Euclidean geometrical volume, being independent
on the matter sector in contrast with the EPS approach, in which the volume is conjugate to the pressure given by the cosmological constant and depends
on the matter content of the theory; (iii) regarding the quantities $E, H, M$ defined in HT, the authors of Ref. \cite{hansen2017universality} make
the following proposal: $M$ is the standard BH mass and $E$ is the {\it horizon curvature energy} (the energy required to warp
spacetime so that it embeds an horizon). Interestingly, $E$ vanishes for planar and toroidal BHs and can be negative for
hyperbolic and higher genus BHs. From a geometric point of view, it has been noted that $E$ is related with the transverse
geometry of the horizon \cite{paranjape2006thermodynamic} and with the generalized Misner--Sharp mass (evaluated at the
horizon), $\mathcal{M}(r_{+})$, which is given by \cite{maeda2008generalized},

\begin{equation}
\label{MisnerHT}
\mathcal{M}(r_{+})=E+P_{\lambda}V.
\end{equation}
For any matter content, it has been shown \cite{cai2008generalized} that $\mathcal{M}$ satisfies the generalized first law \cite{hayward1998unified}. Regarding the enthalpy, for AdS--RN BHs we have that \cite{hansen2017universality} $M=H+Q\phi+2P_{m}V$, where $P_{m}$ stands for the pressure corresponding to the matter sector.
Therefore, only when $P_{m}=0$ we have $H=M=E+P_{\lambda}V$. Finally, we recall that only in vacuum and for $P_{tot}>0$, both the EPS and the HT
approaches yield the same kind of thermodynamic behaviors and phase transitions \cite{hansen2017universality}.
Finally, following \cite{hansen2017universality} it is easy to see that the EoS for AdS--RN BHs is given by
\begin{equation}
\label{eosht}
P_{tot}=\frac{T}{2r_{+}}-\frac{1}{8\pi r_{+}^2},
\end{equation}
where $V$ is given by Eq. (\ref{VolRNAdS}) and
\begin{equation}
P_{tot}=P_{m}+P_{\lambda},
\end{equation}
with
\begin{equation}
\label{PQ1}
P_{m}=T^{r}_{\,\,r}=\frac{Q^2}{8\pi r_{+}^4},
\end{equation}
is the radiation pressure exerted on the horizon due only to the Maxwellian source terms.
Finally, we note that, although Eqs. (\ref{eoseps}) and (\ref{eosht}) coincide,
the thermodynamical behavior which they describe is different, as pointed out in \cite{hansen2017universality}.

\section{Equation of state and $\mathcal{K}$-curvature}
\label{GS}
In this section, a geometric interpretation for the EoS describing an AdS--RN BH is developed. In particular, it is shown
that the mentioned EoS can be derived from the concept of $\mathcal{K}$-curvature developed by Penrose and Rindler \cite{penrose1984spinors}.


Let us consider a spherically symmetric and static geometry describing a BH within GR (including a negative cosmological constant, $\lambda$) coupled with Maxwell electrodynamics.
Using the Schwarzschild ansatz we can write
\begin{equation}
\label{metric}
ds^2=f(r)dt^2-f(r)^{-1}dr^2-r^2 d\Omega^2.
\end{equation}
After choosing the following null tetrad
\\
\begin{eqnarray}
l^{a}&=&(1,f,0,0), \nonumber \\
n^{a}&=&(\frac{1}{2f},-\frac{1}{2},0,0), \nonumber \\
m^{a}&=&(0,0,\frac{1}{\sqrt{2}r},\frac{i\, \textrm{csc} \theta}{\sqrt{2}r})
\end{eqnarray}
\\
with ${\bf l} \cdot {\bf n}=1$ and ${\bf m} \cdot {\bf \bar m}=-1$, with the bar denoting complex conjugation,
the only non--vanishing NP symbols for the static Einstein--Maxwell system are
\\
\begin{eqnarray}
\Psi_{2}&=&C_{pqrs}l^{p} m^{q} \bar m^{r} n^{s}, \nonumber \\
\Phi_{11}&=&-\frac{1}{2}R_{ab}l^{a}n^{b}+ 3 \Lambda, \nonumber \\
\Lambda&=&\frac{R}{24}.
\end{eqnarray}
\\
Specifically, for a metric given by Eq. (\ref{metric}) we get
\\
\begin{eqnarray}
\Psi_{2}&=&-\frac{1}{6 r^2}+\frac{f}{6 r^2}-\frac{f'}{6 r}+\frac{f''}{12},\nonumber \\
\Phi_{11}&=&\frac{1}{4 r^2}-\frac{f}{4 r^2}+\frac{f''}{8}\nonumber, \\
\Lambda&=&\frac{1}{12 r^2}-\frac{f}{12 r^2}-\frac{f'}{6 r}-\frac{f''}{24}.
\end{eqnarray}
\\
At this point, a couple of comments are in order. First, note that, after introducing the Misner-Sharp mass,
$\mathcal{M}(r)$, as
\begin{equation}
f(r)=1-\frac{2\mathcal{M}(r)}{r},
\end{equation}
the following relation can be obtained:
\begin{equation}
\label{Misner}
\mathcal{M}=(\Phi_{11}-\Psi_{2}+\Lambda)r^3,
\end{equation}
whose validity extends beyond our metric form (\ref{metric}), as discussed below in Section \ref{Discussion}.

Second, the Komar energy,
\begin{equation}
E_{K}=-\frac{1}{8\pi}\int_{S^{2}(r)}dS_{\mu\nu}\nabla^{\mu}\xi^{\nu},
\end{equation}
where $dS_{\mu\nu}$ denotes the surface element on $S^{2}(r)$
and $\xi^{\mu}=(1,0,0,0)$ is a timelike Killing vector, can be written, as checked by explicit calculation, as
\begin{equation}
\label{Komar}
E_{K}=-(2\Lambda+\Psi_{2})r^{3}.
\end{equation}
Therefore, using Eqs. (\ref{Misner}) and (\ref{Komar}), we obtain a relation between the Komar energy and the Misner mass
\begin{center}
\begin{equation}
\label{Komar-Misner}
E_{K}=-(3\Lambda+\Phi_{11})r^3+\mathcal{M}.
\end{equation}
\end{center}
Let us now consider the so-called holographic energy equipartition \cite{padmanabhan2010thermodynamical} which, for a static spacetime, reads
\begin{equation}
\label{Padmaequi}
E_{K}\equiv \int_{\mathcal{V}} d^{3}x\sqrt{h}\rho_{K}\,\hat{=}\,\frac{1}{2}\int_{\partial \mathcal{V}} \frac{d^{2}x\sqrt{\sigma}}{l_{p}^{2}}
T_{\mathrm{loc}},
\end{equation}
where $h$ and $\sigma$ are the induced metrics defined on $\mathcal{V}$ and $\partial \mathcal{V}$, respectively, $T_{\mathrm{loc}}$ stands for the local Hawking temperature measured by an observer at rest in this spacetime and $\rho_{K}$ is defined as a Komar energy--density. The symbol $\hat{=}$ is used to specify that the equality is only valid at the horizon, $r=r_+$, and it will be used with such meaning from now on. Following \cite{padmanabhan2010thermodynamical} we can attribute $\Delta N=d^{2}x\sqrt{\sigma}l_{p}^{-2}$ microscopic degrees of freedom to an area element $\Delta A=d^{2}x\sqrt{\sigma}$.

Even more, Eq. (\ref{Padmaequi}) can be written  on the horizon, $r_{+}$, as
\begin{equation}
\label{equip}
E_{K}\,\hat{=}\,\frac{1}{2}r_{+}^2 f'(r_{+})= \frac{1}{2}N T,
\end{equation}
where $T=\frac{f'(r_{+})}{4\pi}$ and $N=A=4\pi r^{2}_{+}$.

In the AdS-RN case, by taking the trace of Einstein equations we get
\begin{equation}
\Lambda=-\frac{1}{2l^2}.
\end{equation}
Therefore, Eq. (\ref{Komar-Misner}) can be written as
\begin{equation}
\label{modified}
E_{K}=\mathcal{M}-\left(\Phi_{11}-\frac{3}{2 l^2}\right)r^3.
\end{equation}
Thus, if the pressure $P_{\lambda}$ is identified with $\frac{3}{8\pi l^2}$, which is the essential assumption of the EPS formalism briefly summarized in the previous section, and after considering that $E_{K}\,\hat{=}\,\frac{1}{2}N T=2 T S$, where $S$ is the
entropy of the BH, Eq. (\ref{modified}) can be written as
\begin{equation}
\label{general}
\mathcal{M} \,\hat{=}\, 2\, T\, S-3(\omega+P_{\lambda})V,
\end{equation}
where $\omega$ is, for the moment, a pressure contribution defined as
\begin{equation}
\label{omega}
\omega=-\frac{\Phi_{11}}{4\pi},
\end{equation}
\\
and $V=\frac{4 \pi r^{3}_{+}}{3}$ is the areal volume, which corresponds, in the EPS approach, to the thermodynamic volume for AdS--RN BHs, as previously stated.

Up to this point, some comments are in order.
First, note that the standard Smarr relation for AdS-RN BHs is recovered. In the uncharged case we get
\\
\begin{eqnarray}
\Psi_{2}&\,\hat{=}\,&-\frac{M}{r_{+}^3}, \nonumber \\
\Phi_{11}&=&0, \nonumber\\
\Lambda&=&-\frac{1}{2 l^2}, \nonumber \\
\mathcal{M}&\,\hat{=}\,&M-\frac{r_{+}^3}{2l^{2}}, \nonumber \\
M&\,\hat{=}\,&2 T S-2 P V.
\end{eqnarray}
\\
Second, in the charged (Maxwell) case we have
\\
\begin{eqnarray}
\Psi_{2}&\,\hat{=}\,&\frac{Q^2-M r_{+}}{r_{+}^4}, \nonumber \\
\Phi_{11}&\,\hat{=}\,&\frac{Q^2}{2 r_{+}^4}, \nonumber \\
\Lambda&=&-\frac{1}{2 l^2}, \nonumber \\
\mathcal{M}&\,\hat{=}\,&M-\frac{Q^2}{2r_{+}}-\frac{1}{2 l^2}r_{+}^3, \nonumber \\
M&\,\hat{=}\,&2 T S+ \phi Q -2 P V,
\end{eqnarray}
\\
where $\phi$ is taken to be the electric potential at the horizon. And third, Eq. (\ref{general}) can be written as the VdW EoS given by Eq. (\ref{VdWRN}) or Eq. (\ref{eoseps}).

Now, we will connect the previous EoS with the geometric quantities used in the SC formalism, which is one of our main results. Penrose and Rindler \cite{penrose1984spinors} introduce the (complex) $\mathcal{K}$-curvature of any spacelike two--surface in spacetime:
\begin{equation}
\label{K}
\mathcal{K}=-\sigma \lambda -\Psi_{2}-\tilde \rho \mu +\Phi_{11} + \Lambda,
\end{equation}
where $\sigma=m^am^b\nabla_bl_a$, $\lambda=\bar{m}^a\bar{m}^b\nabla_bn_a$, $\tilde \rho=m^a\bar{m}^b\nabla_b l_a$ and $\mu=\bar{m}^a m^b\nabla_b n_a$, are spin coefficients related to the expansion and shear of the null congruences with tangent vectors
$l^a$ and $n^a$. Even more, they show \cite{penrose1984spinors} that
\begin{equation}
\label{gauss}
\mathcal{K}+\mathcal{ \bar K}=k_{g},
\end{equation}
where $k_{g}$ is the Gaussian curvature of the considered two--surface and $\mathcal{ \bar K}$ is the complex conjugate of $\mathcal{K}$. For an AdS--RN BH horizon (generated by a shear-- and expansion--free null congruence, where $\tilde{\rho}=0=\sigma=\lambda$), Eq. (\ref{gauss}) reads
\begin{equation}
\label{Delta}
\frac{k_{g}}{2}+ \Psi_{2}-\Phi_{11}-\Lambda=\frac{f}{2r^{2}}\,\hat{=}\,0,
\end{equation}
where the first equality is provided by explicit computation and proves that theorem (\ref{gauss}) is, when evaluated at a horizon, equivalent to the vanishing of the metric function $f$.
Once the corresponding values for the NP scalars
for the AdS--RN solution are computed, we write the mass as a function of the temperature using
Eqs. (\ref{Mass}) and (\ref{TempRNAdS}), obtaining
\begin{equation}
\label{mass}
M=2\frac{r_{+}^3}{l^2}+r_{+}-2\pi r_{+}^2 T.
\end{equation}
Then, introducing Eq. (\ref{mass}) in Eq. (\ref{gauss}) we obtain Eq. (\ref{eoseps}), which we remind the reader is given by
\begin{equation}
P_{\lambda}=\frac{\bar N T}{V}-\frac{1}{2\pi}\frac{\bar N^2}{V^2}+\frac{2 Q^2}{\pi}\frac{\bar N^{4}}{V^{4}}.
\end{equation}
In this equation of state, the following identification can be performed
\begin{eqnarray}
-\frac{3}{4\pi}\Lambda&=&P _{\lambda},\nonumber \\
-\frac{1}{4\pi}\left(\Psi_{2}+2 \Lambda\right)&=&\frac{\bar N T}{V}, \nonumber \\
\frac{1}{4\pi}\mathcal{K}=\frac{1}{4\pi}\frac{1}{2}k_{g}&=&\frac{1}{2\pi}\frac{\bar N^2}{V^2}, \nonumber \\
\frac{1}{4\pi}\Phi_{11}&=&\frac{2 Q^2}{\pi}\frac{\bar N^{4}}{V^{4}},\label{EPSidentification}
\end{eqnarray}
\\
which can be taken as the geometric splitting of the EoS for AdS--RN BHs in the EPS setting. It is important to note that this identification is not unique, but it depends on the thermodynamical framework in which one is working. For example, in the HT approach we have that the pressure is defined in terms of $T^r_{\,\,r}$, as reviewed above; in this context, the corresponding identification is given by
\begin{eqnarray}
-\frac{1}{4\pi}\left(\Phi_{11}+3\Lambda\right)&=&P_{tot},\nonumber \\
-\frac{1}{4\pi}\left(\Psi_{2}+2 \Lambda\right)&=&\frac{\bar N T}{V}, \nonumber \\
\frac{1}{4\pi}\mathcal{K}=\frac{1}{4\pi}\frac{1}{2}k_{g}&=&\frac{1}{2\pi}\frac{\bar N^2}{V^2}.\label{HTidentification}
\end{eqnarray}
The main difference, as discussed in \cite{hansen2017universality}, lies in the fact that the HT pressure includes contributions from all matter sources; thus, we can not split NP scalars as associated to the cosmological constant and the radiation pressure from a thermodynamic point of view. In addition, we must remark that the independence of the thermodynamic description with respect to the explicit matter sources that characterizes HT is recovered in these latter results; therefore, the extension of these identifications to situations with additional sources such as scalar fields is straightforward.
\\
Summarizing, the proposed identifications let us conclude that the equation of state given by Eq. (\ref{eoseps}) is nothing but the theorem expressed by Eq. (\ref{gauss}) linking
the $\mathcal{K}$--curvature with the Gaussian curvature of the horizon of an AdS-RN BH. Specifically for the EPS case, the NP scalars are involved in the following way: $\Lambda$
corresponds to the pressure, the Komar density corresponds to the kinetic term of the ideal gas, the complex curvature $\mathcal{K}$ corresponds to
the VdW ``interaction term" and $\Phi_{11}$ corresponds to the second virial term which describes the Maxwellian radiation pressure exerted on the horizon. In the case of HT, there is no second virial term since it has to be regarded as part of the thermodynamical pressure. This fact leads to striking differences between the two approaches with respect to the phase behavior and the interpretation of the results, that are analogous to the results of \cite{hansen2017universality}. In addition to these points, the correspondence that we obtained permits, in principle, to see how the geometric information encoded in the NP scalars ``emerges" from statistical mechanical models such as the one proposed in Ref. \cite{vargas2018sads}. The corresponding development for the interaction term was already discussed in detail in that reference and for the Komar energy in \cite{padmanabhan2010equipartition}.

\section{Pressure/energy density conditions at the horizon}
\label{GSGEM}

From the set of non-vanishing NP scalars for our static spherically symmetric case, $\{\Lambda,\Psi_{2},\Phi_{11}\}$, and their combination in terms of $\mathcal{K}$, we have identified $\Lambda$ and $\Phi_{11}$ with pressure terms, the first corresponding to the cosmological pressure and the second one to the Maxwellian radiation pressure exerted on the horizon. In addition, we established that these identifications are not unique but depend on the thermodynamic framework in which the BH are described. Our purpose in this Section is to take this idea further and provide another framework in which pressure-like interpretations are considered for the whole set of NP scalars, with the consequence that it is possible to define the event horizon in terms of a sum of these pressures. By virtue of the respective equations of state for each element, such a sum can also be understood as a condition on the total energy density, which is interesting and deserves to be discussed.

Let us now motivate our subsequent discussion by fixing our attention in the second equality of Eq. (\ref{Delta}), which, remembering that this equation is equivalent to the vanishing of the metric function at the horizon, reads
\begin{equation}
\frac{1}{2r_{+}^2}-\frac{M}{r_{+}^3}+\frac{Q^2}{2r_{+}^4}+\frac{1}{2 l^2}=0,
\end{equation}
or
\begin{equation}
\label{pressures1}
\left(\frac{3}{4\pi}\frac{1}{2r_{+}^2}\right)-\left(\frac{M}{V}\right)+3\left(\frac{1}{4\pi}\frac{Q^2}{2r_{+}^4}\right)
+\left(\frac{3}{8\pi l^2}\right)=0.
\end{equation}
Notice that, since we are working with units such that $4\pi \epsilon_{0}=1$, the third term of Eq. (\ref{pressures1})
represents the radiation pressure exerted on the horizon, which constitutes the matter contribution, $P_{m}$, relevant for this case under the HT approach, as previously stated. That is,
\begin{equation}
\label{PQ2}
P_{m}=\frac{1}{4\pi}\frac{Q^2}{2r_{+}^4}.
\end{equation}
If we naively define an
energy density, $\rho_{M}$, as $M/V$, then Eq. (\ref{pressures1}) reads
\begin{equation}
\label{pressures2}
\left(\frac{3}{4\pi}\frac{1}{2r_{+}^2}\right)-\rho_{M}+3 P_{m}+P_{\lambda}=0,
\end{equation}
or
\begin{equation}
\label{pressures3}
-\left(\rho_{M}+\rho_{\lambda}\right)+3 (P_{m}+P_{\sigma})=0,
\end{equation}
\\
where the energy density associated with the cosmological constant, $\rho_{\lambda}=-P_{\lambda}$ has been introduced in
order to facilitate the interpretation of Eq. (\ref{pressures3}) and we have introduced a
{\it horizon curvature pressure} on the horizon:
\begin{equation}
P_{\sigma} = \frac{k_{g}}{8\pi}.
\end{equation}
This pressure was defined originally in \cite{hansen2017universality} for general horizons and can be associated to their curvature since its sign depends on the sign of the 2-curvature of such surfaces, and also because it vanishes for planar horizons. It is useful to note, given the upcoming discussion, that a local equation of state can be constructed for this pressure by considering its relation with the {\it horizon curvature energy} mentioned above. Namely, we can define a horizon curvature energy density for this energy, which is given by $E=\frac{r_+}{2}$, as
\begin{equation}\label{rhosigma}
\rho_\sigma\equiv\frac{E}{V_t}=\frac{3}{4\pi}\frac{1}{2r_+^2},
\end{equation}
with $V_t$ the Euclidean volume considered in HT. From this expression, explicit calculation leads to the following equation of state for the horizon curvature variables
\begin{equation}\label{sigmaeos}
P_\sigma=\frac13\rho_\sigma.
\end{equation}
At this point, one could be tempted to read Eq. (\ref{pressures3}) as some kind of thermodynamical condition on the horizon but,
in order to do that, an object defining ``a pressure related to $M$" has to be introduced. In fact, this is largely artificial
since both $M$ and $Q$ contribute to the Riemann curvature.
In this line of thought, what makes more sense is to split the Riemann curvature as usual in its traceless and matter parts
using the Weyl and Ricci curvatures. Following this argument, here we will see that Eq. (\ref{pressures3}) and, therefore, Eq. (\ref{K}), can be
written in terms of the gravitational energy-momentum by using the
SQBR tensor \cite{bonilla1997some}.

For a generic spacetime, the Bel-Robinson tensor is defined as \cite{bel1958definition,bel1959introduction,bel2000radiation}
\begin{equation}
T_{abcd}=C_{aecf}C^{\,\,e\,\,f}_{b\,\,d\,}+ \,^{\ast}C_{aecf} \,^{\ast}C^{\,\,e\,\,f}_{b\,\,d},
\end{equation}\\
where $C_{aecf}$ corresponds to the Weyl tensor and $*$ denotes the Hodge dual in four dimensions. The BR tensor is completely symmetric, traceless  and covariantly conserved in vacuum. Given a generic timelike congruence, $u^{a}$, a
super--energy density can be defined as $W=T_{abcd}u^{a}u^{b}u^{c}u^{d}$. Even more, as the BR tensor has dimensions of $[L]^{-4}$,
a properly defined square root could account for a possible definition of the energy--momentum tensor for free gravitational
fields \cite{bonilla1997some}. The SQBR tensor, $t_{ab}$, is a symmetric, two--index tensor which is solution of \cite{schouten2013ricci}
\begin{widetext}
\begin{equation}
\label{Schouten}
T_{abcd}=t_{(ab}t_{cd)}-\frac{1}{2}t^{\,e}_{e}t_{(ab}g_{cd)}+\frac{1}{24}\left(t_{ef}t^{ef}+\frac{1}{2}(t^{\,e}_{e})^2 \right)
g_{(ab}g_{cd)}.
\end{equation}
\end{widetext}
In fact, $t_{ab}+H g_{ab}$, where $H$ is an arbitrary function, is also a solution of Eq. (\ref{Schouten}).

Exploiting that the solutions we are interested in are classified within the Petrov-D type, then $t_{ab}$ can be written as \cite{bonilla1997some}
\begin{equation}
t_{ab}=2 \,c\, |\Psi_{2}|\left(l_{(a}n_{b)}+m_{(a}\bar m_{b)} \right)+ H g_{ab},
\end{equation}
where $c$ is an arbitrary constant introduced in order to compare with different possible conventions as in
Ref. \cite{acquaviva2018gravitational}. Different choices for the arbitrary function $H$ have appeared in the literature, basically
depending on the vanishing of the covariant divergence of $t_{ab}$, $u_{a}\nabla_{b}t^{ab}$. On one hand, the form of $H$ consistent
with covariant conservation has been considered in Refs. \cite{clifton2013gravitational,acquaviva2015constructing} and, on the other hand, $H=0$ has been chosen
in Ref. \cite{acquaviva2018gravitational} in order to secure a traceless $t_{ab}$ and, therefore, a massless carrier of the gravitational field.

A fluid--like interpretation of the SQBR tensor can be specified once a timelike congruence $u^{a}$, with $u^{a}u_{a}=1$, has been chosen. Such congruence is interpreted as a family of observers carrying a four--velocity $u^{a}$. If the associated orthogonal projector is $h_{ab}=g_{ab}-u_{a}u_{b}$, any tensor (in particular the SQBR tensor) can be decomposed as
\begin{equation}
t_{ab}=\mu_{g}u_{a}u_{b}-2q_{(a}u_{b)}+P_{g}h_{ab}+\pi_{ab}.
\end{equation}
For the SQBR tensor, $\mu_{g},q_{a},P_{g}$ and $\pi_{ab}$ can be taken to be the energy density, heat flux, isotropic and anisotropic pressures,
respectively, associated to gravitation and measured by the observer $u^{a}$. Note that these identifications are strongly linked with the identification of the SQBR tensor as the object that describes the energy and momentum of gravitation.

In the frame adapted to the principal null directions (the comoving frame), and choosing $H=0$, these thermodynamic quantities read
\cite{acquaviva2018gravitational}
\\
\begin{eqnarray}
\mu_{g}&=&c \, |\Psi_{2}|, \nonumber \\
P_{g}&=&-\frac{c}{3}|\Psi_{2}|,\nonumber \\
q_{a}&=&0, \nonumber \\
\pi_{ab}&=&\frac{2 \,c}{3}\left(x_{a}x_{b}+y_{a}y_{b}-2 z_{a}z_{b} \right)\label{SQBRidentification}
\end{eqnarray}
\\
where the triad $\{x^{a},y^{b},z^{c}\}$ is a set of orthonormal basis vectors (see \cite{acquaviva2018gravitational} for details).
Interestingly, the EoS
\begin{equation}
\label{pgrav}
P_{g}=-\frac{1}{3}\mu_{g}
\end{equation}
is also valid for type N spacetimes, being invariant under general Lorentz transformations \cite{acquaviva2018gravitational}.
On the contrary, if $u_{a}\nabla_{b}t^{ab}=0$ is taken as the main constraint of the SQBR tensor, then $H \sim -\Psi_{2}$
and $P_{g}=0$ \cite{clifton2013gravitational}.

Let us now fix our attention in the first equality of Eq. (\ref{Delta}), which reads
\begin{equation}
\label{kg}
\frac{k_{g}}{2}+ \Psi_{2}-\Phi_{11}-\Lambda=0.
\end{equation}
If we choose $H=0$, and consider that $\Psi_2$ is negative for well behaved AdS-RN BHs, then Eq. (\ref{kg}) can be written as
\begin{equation}
\label{pressures4}
-\left(\frac{3}{4\pi c}\mu_{g}+\rho_{\lambda}\right)+3 (P_{\sigma}-P_{m})=0.
\end{equation}
or
\begin{equation}
\label{pressures5}
P_{\lambda}-3(P_{m}-P_{\sigma}-\frac{3}{4\pi c}P_{g})=0,
\end{equation}
%

which can be interpreted as a pressure constraint that {\it defines} the horizon.
Note that the combination of signs which appears in Eq. (\ref{pressures5}) is determined by those of Eq. (\ref{K}).
Finally, we note that Eq. (\ref{pressures5}) clearly shows that the curvature pressure at (or defining the) horizons can be decomposed on matter
(including cosmological) and gravitational components.
\\
\\
We would like to stress again that Eq. (\ref{pressures5}) is completely equivalent to Eq. (\ref{Delta}), which can be used to define the horizon.
Then, $r=r_{+}$ is a horizon for the AdS-RN BH when
\begin{equation}
\sum_{i=g,m,\lambda,\sigma}\alpha_i P_{i}(r_{+})=0,
\end{equation}
where the corresponding pressures have been defined above and $\alpha_i$ are the respective numerical factors. It is important to note that, by virtue of the EoS for the different components, it is possible to obtain a energy density version of (\ref{pressures5}) which has an interesting form. Namely:
\begin{equation}\label{energydensities1}
\frac{3}{4\pi c}\mu_g+\rho_\lambda-\rho_\sigma+\rho_m=0.
\end{equation}
As discussed above, $c$ is introduced as a dimensionless numerical constant whose introduction is motivated mainly by the comparison between conventions; however, we see that its effects are non-trivial in these equations. Specifically, if we define a (local) total energy density for the horizon as $\rho_{net}=\rho_m+\rho_\lambda+\rho_\sigma+\mu_g$, it follows from (\ref{energydensities1}) that
\begin{equation}\label{total-energy-density}
\rho_{net}=2\rho_\sigma+\left(1-\frac{3}{4\pi c}\right)\mu_g=|\Psi_2|\left(c-\frac{3}{4\pi}\right).
\end{equation}
Thus, it is evident that $c$ has an important effect in the value of the total energy density. As an example, we can consider the case $c=\frac{3}{4\pi}$, where Eqs. (\ref{pressures4}), (\ref{pressures5}), and (\ref{energydensities1}) take the form,
\begin{equation}
\label{pressures6}
-\left(\mu_{g}+\rho_{\lambda}\right)+3 (P_{\sigma}-P_{m})=0,
\end{equation}
\begin{equation}
\label{pressures7}
P_{\lambda}-3(P_{m}-P_{\sigma}-P_{g})=0,
\end{equation}
and
\begin{equation}\label{energydensities2}
\mu_g+\rho_\lambda+\rho_m=\rho_\sigma.
\end{equation}
This equation states, under the suppositions described, that the horizon could be defined as a spherical surface whose horizon energy density equals the remaining energy content, including gravitation. At this point, it is important to consider whether additional conditions exist which could lead us to expect some value for $\mu_g$, since this condition amounts to a physical argument to choose a value for $c$. This is an interesting point that requires further research. Nevertheless, we must remark that our findings show that, in any case, an horizon can be defined through a thermodynamical condition on the total energy in terms of the energies of the different elements that compose the system, this statement being also valid for the pressure condition (\ref{pressures5}). In particular, for this condition, our result certainly can not be interpreted a priori as an equilibrium condition, as evident from the coefficients in the pressure equation; instead, these coefficients are more general and could be associated to a condition on decoupled substances in contact.

In a broader sense, we can conclude that there is another way to define horizons, from a thermodynamical point of view, in addition to the usual considerations about the Einstein equations as either realizations of the first law (HT) or EoS, and the interpretation of geometrical charges as thermodynamical potentials such as enthalpy (EPS); namely, as surfaces where energy densities or pressures of the constituents must satisfy a specific condition. It is expected that this rationale extends in static spherically symmetric metrics beyond our AdS-RN setting, by identifying the appropriate pressure and energy density terms for the additional sources in the context of hairy black holes, for example. From our point of view, microscopical models could be introduced to provide an statistical foundation for these conditions in the context of emergent gravity, although we consider that the absence of a temperature term leaves a greater range of possibilities open for these models than in the other approaches to BH thermodynamics, which is an issue when trying to restrict the microscopic models from which gravitation could emerge.

\section{Discussion}
\label{Discussion}
In this section we make some remarks with respect to the obtained results and discuss the different connections with previous work in the literature. In particular, we will focus in two aspects: the relation of our EoS in terms of NP scalars with the general laws of BH dynamics first laid out by Hayward in the language of SC \cite{hayward1994spin}. Finally, we give some remarks with respect to the thermodynamic identifications we found, the differences between different approaches such as HT and EPS, and give some conclusions.

First of all, it is necessary to discuss our results in the context of previous work by Hayward \cite{hayward1994spin}. As we pointed out before, this work states the laws of BH dynamics in terms of SC. These laws have an essentially geometric character, so it is interesting to remark the similarities and differences with BH thermodynamics in terms of EoS, beyond the lack of prescriptions for the identification of pressures in such geometric setting that we mentioned above. Hayward's work is based on the study of the evolution of spacetime along null congruences; in fact, in previous work \cite{hayward1993dual}, Hayward also constructs the so-called dual-null dynamics of the gravitational field along this guideline. For our purposes, we summarize here some results of these works that we consider relevant for our discussion. In equation (\ref{Misner}) we associated the Misner-Sharp mass to a certain combination of NP scalars; in fact, this relation is a particular case of the Equation (21) of \cite{hayward1994spin}, which states that
\begin{equation}\label{Misner-Hayward}
\mathcal{M}=\frac{1}{2\pi}\sqrt{\frac{A}{16\pi}}\int_S*\frac{\mathcal{K}+\tilde{\rho}\mu-\sigma\lambda+\tau\tau'}{\chi\bar{\chi}},
\end{equation}
where $\mathcal{M}$ is to be understood, in this context, as the double-null Hamiltonian which reduces to Misner-Sharp mass in the spherically symmetric case, and $\tau=m^a n^b\nabla_b l_a$, $\tau'=\bar{m}^a l^b \nabla_b n_a$ are spin-coefficients related to the twist of the null congruences, and vanish for the spherically symmetric spacetime that we are considering. $S$ is a compact 2-surface with area $A$ and area form $*1$, where the Hogde dual is understood for forms defined on $S$. $\chi$ is fixed from the normalization of the spin basis, in this case $\chi\bar\chi=l^a n_a=1$. After replacing the definition of the $\mathcal{K}$-curvature, it is found that
\begin{equation}\label{Misner-Sharp-general}
\mathcal{M}=\frac{1}{2\pi}\sqrt{\frac{A}{16\pi}}\int_S*\frac{-\Psi_2+\Phi_{11}+\Lambda+\tau\tau'}{\chi\bar{\chi}}.
\end{equation}
When we consider $S$ to be a constant radius 2-surface in a spherically symmetric spacetime, Equation (\ref{Misner}) is recovered. Thus, upon identification of the NP scalars with pressures we have the corresponding expression (\ref{Misner-Sharp-general}) that applies to general spacetimes. It remains to be seen if a thermodynamical interpretation can be provided for $\tau$ and $\tau'$ in stationary spacetimes, which we consider necessary for the analysis of rotating spacetimes within our approach.

More connections can be identified between our results and the approach of \cite{hayward1994spin}. In fact, one could ask whether thermodynamical approaches such as EPS or HT can be reconstructed in the language of SC and dual-null dynamics. This is indeed the case for HT. As reviewed above, this framework is constructed from the radial Einstein equation in a spherically symmetric spacetime, therefore it is to be expected that one of the equations describing the evolution of SC can equivalently fulfill such role. This equation is the {\it cross-focusing equation}, Eq. (17) of \cite{hayward1994spin}:
\begin{equation}\label{cross-focusing}
\text{\textthorn}'\tilde{\rho}-\eth'\tau=\tilde{\rho}\mu+\sigma\lambda-\tau\bar{\tau}-\Psi_2-2\Lambda,
\end{equation}
where \textthorn$'$ can be understood as a scaling-invariant version of the directional derivative along the null vector $n^a$ \footnote{In fact, for this symmetry it coincides with such directional derivative}, whereas $\eth'$ is the corresponding object for one of the spacelike directions on the sphere, $\bar{m}^a$ (see \cite{penrose1984spinors} for more details). In the case of a horizon in a spherically symmetric spacetime, the vanishing of the spin coefficients mentioned above implies that this equation reduces to
\begin{equation}\label{cross-focusing-ssym}
\text{\textthorn}'\tilde{\rho}\,\hat{=}\,-\Psi_2-2\Lambda.
\end{equation}
Hayward \cite{hayward1994spin} argues that the trapping gravity, which in the dynamical BH laws is analogous to the temperature, is defined in terms of \textthorn$'\tilde{\rho}$; furthermore, explicit calculation for the metric (\ref{metric}) shows that this object is equal to $f'(r_+)/2r_+$ on the horizon, and then proportional to the BH temperature. From this fact, Eq. (\ref{cross-focusing-ssym}) can be taken as the starting point for HT by considering the definition of $\mathcal{K}$, Eq. (\ref{K}), evaluated at the horizon:
\begin{equation}\label{K-horizon}
\mathcal{K}\,\hat{=}\,-\Psi_2+\Phi_{11}+\Lambda.
\end{equation}
This equation, together with (\ref{cross-focusing-ssym}), implies that
\begin{equation}\label{np-ht}
\text{\textthorn}'\tilde{\rho}\,\hat{=}\,\mathcal{K}-\Phi_{11}-3\Lambda.
\end{equation}
The contributions associated to the HT EoS can be recognized in this equation. By Penrose's theorem (\ref{gauss}), we know that $\mathcal{K}=k_g/2$; in addition, as explained in \cite{hansen2017universality}, internal energy in HT is proportional to the curvature of the horizon, $k_g$, so we have that the $\mathcal{K}$ term in this equation provides the internal energy term of the integrated first law in HT. Furthermore, we identified before, in Eq. (\ref{HTidentification}), that $-(\Phi_{11}+3\Lambda)$ is the pressure for AdS-RN BH within the HT approach according to the cosmological constant and radiation pressures; therefore, we obtain that this identification for the pressure in terms of $\Phi_{11}$ and $\Lambda$ applies to any spherically symmetric BH. Interestingly, this definition provides a connection with energy conditions in this context since $\Phi_{11}+3\Lambda\ge0$ is one consequence of the dominant energy condition, so we have that the HT pressure in a spacetime that obeys this energy condition must be negative. This identification provides a way to cast BH dynamical laws in thermodynamical terms; for example, Hayward topology theorem \cite{hayward1994spin} implies, in these terms, that a BH with negative pressure, necessarily must have positive curvature energy (equivalent to $k_g>0$). Since many results in \cite{hayward1994spin} are a consequence of the dominant energy condition, thermodynamical statements for HT pressure in spherically symmetric spacetimes can be readily obtained.

Although we found interesting connections between Hayward's BH dynamical laws and the BH EoS in terms of NP scalars, it is important to note that there are differences; in particular, regarding the laws with an explicit dynamical component. For example, Hayward's first law, Eq. (10) of \cite{hayward1994spin}, establishes that the area form of the horizon, $*1$, changes proportionally to the square root of the NP symbol $\Phi_{00}$, which is zero for the type-D spacetimes we are considering. The character of this law is fundamentally different from the first law of BH thermodynamics in the HT and EPS approaches; in the first case, the variation of area (entropy) is connected to {\it virtual} displacements of the horizon, whereas in the second one, we compare asymptotically AdS solutions with slightly different parameters, as mentioned before. In fact, this difference has been pointed out before and its importance lies in that horizon area and entropy are not connected in dynamical spacetimes in a straightforward way. In spite of this issue, it is remarkable that the dynamical first law can give us lessons that are relevant for the BH thermodynamic approaches we are using. Namely, that different Petrov type-D spherically symmetric static spacetimes can not be connected dynamically through states of the same type, we must have a non vanishing $\Phi_{00}$ to be able to generate evolution from one to the other; this requirement is to be expected since qualitatively we can imagine that the feeding of a BH with matter to increase its area implies both a flux of energy-momentum for the feeding process and the emission of gravitational waves produced while the BH reaches it final state. This reasoning gives us an idea of the high level of idealization required to describe BH processes in terms of quasistatic trajectories in a space of states and, in addition, is important in the context of approaches to thermodynamic equilibrium for BHs.

In the case of the relation of the EPS approach and the NP symbols, we must remark that the generalization of our identification to other spacetimes is not straightforward. In contrast with HT, where the universality of the construction allows for a definition of thermodynamics irrespective of the concrete form of the metric, the EPS formulation of BH thermodynamics depends strongly on the matter sources. This can be seen as a trade-off for a richer phase structure than the one present in the HT case. For example, in our study of AdS-RN BHs we associated the EPS pressure, defined in terms of the cosmological constant, to the $\Lambda$ scalar, which is proportional to the Ricci scalar; however, in other situations could be contributions to this scalar which are not associated with the cosmological constant, as in the case of massive electrodynamics. Thus, the connection of EPS thermodynamical quantities to NP scalars proceeds on a case by case basis. In a broader setting, it is interesting to note that the character of the thermodynamical descriptions is very different in our case and in EPS. We study the behavior of certain scalars {\it at the horizon}, obtaining that local geometrical relations can be interpreted as thermodynamical relations measured by certain observers, which put us closer to the spirit of HT, and even Hayward's work \cite{hayward1994spin}; on the other hand, EPS considers laws for the whole spacetime since, as reviewed before, it compares solutions with slight differences in their parameters. The horizon relations enter in this context as a change of variable from $M$ to $r_+$, but this should not make us to lose sight of the structure of the approach. Thus, to map this structure in terms of NP symbols, defined locally, is not easy in general terms, even as the explicit EoS can be readily connected. These issues are far more important when devising applications, since the possible interpretations of the results depend on them.

One last aspect we would like to discuss is the freedom in the choice of thermodynamic variables that our results suggest. Summarizing, we have seen that there are different ways to assign a thermodynamical role to the geometric objects that we are interested in, the NP scalars and $\mathcal{K}$; namely, the equations (\ref{EPSidentification}) and (\ref{HTidentification}), together with the associations with pressures and energy densities based on the SQBR tensor, Eq. (\ref{SQBRidentification}).

With these connections in mind, one could ask if there are some additional criteria that could help to filter out some alternatives. Naturally, this is the case. One criterion is given by the same reasoning that led historically from the analogies in BH physics to the recognition of the true thermodynamical nature of BH: the existence of processes, such as Hawking radiation, that implement physically the thermodynamic roles that we assign to geometric quantities. For example, in our construction of gravitational pressures via the SQBR tensor it is interesting to study what do mean the identifications that we performed for pressure and energy density of the gravitational field in terms of $\Psi_2$. One important issue with the usual thermodynamical reasoning is that the backreaction of the metric to Hawking radiation is neglected, perhaps the emitted particles have an associated pressure that is related with our definition, although this is only a conjecture that must be studied in depth. In any case, proposals that describe physically the mechanisms underlying pressures and/or energy densities could be valuable; however, the physical implementation of these ideas is framed in the problem of the interpretation of proposals for the energy of the gravitational field, such as the SQBR tensor, and as such, it remains an open problem where consensus has not been reached yet.

Another criterion that could help to obtain insights for the identification of the NP scalars in terms of thermodynamic variables could be the relation with approaches that consider gravity as an emergent phenomena. For example, in the EPS approach we can consider a connection of the asymptotically AdS gravitational system with a field theory via the AdS/CFT correspondence; in this context, thermodynamical variables such as the pressure $P_\lambda$ have a connection with quantities of the theory. In the particular case of the pressure, it has been argued that this variable corresponds to the number of flavors in the CFT, and that thermodynamical processes where pressure varies can be mapped to renormalization group transformations in the CFT \cite{johnson2014holographic}. Another possibility could be a corpuscular model such as the proposal in \cite{vargas2018sads}. We must recognize that we do not know yet what model could provide a microscopical basis for gravity, but, in general, if the underlying theory is able to describe processes with a thermodynamic equivalent in the spacetime under study, then it constitutes an useful guideline for a true identification. However, there is still much work to be done in this context to obtain an answer.

\section{Final Remarks}

In this work we have investigated the connections between Newman-Penrose scalars for spherically symmetric spacetimes and the equations of state for asymptotically Anti-de Sitter Reissner-Nordstr\"om black holes in the context of Horizon Thermodynamics and Extended Phase Space approaches to black hole thermodynamics. In particular, we have shown that the Penrose-Rindler $\mathcal{K}$--curvature corresponds to the generalized Misner mass density associated with the areal volume of the horizon for the studied spacetime, and concluded that this is a particular case of the relation between Newman-Penrose scalars and spin-coefficients with the dual-null Hamiltonian introduced by Hayward \cite{hayward1994spin}. Also, a geometric splitting is proposed for the equations of state of Horizon Thermodynamics and Extended Phase Space in terms of the non-vanishing Newman-Penrose scalars which define the $\mathcal{K}$-curvature at the horizon. This result provides prescriptions for the identification of pressures among the Newman-Penrose scalars, which is not straightforward in a purely geometric approach. Finally, we arrived to conditions for the pressures or energy densities at (or defining) the horizon, that have been derived by introducing the square root of the Bel-Robinson tensor and a gravitational pressure related to it; such conditions can be thought as thermodynamic definitions of the horizons for the kind of black holes here considered. We also discussed the relation between these results and previous works in this field, identifying directions for future work and open important questions.

Our results allow for a description of black hole thermodynamics in terms of Newman-Penrose scalars, which can be readily linked with previous findings such as the dynamical laws for black holes \cite{hayward1994spin}, and therefore can be useful to provide robust interpretations for thermodynamic developments in the context of Horizon Thermodynamics and Extended Phase Space approaches. In addition, our horizon definitions in terms of pressures and energy densities are interesting since they can be thought of as emergent relations, which allow for a broader set of possibilities regarding models that seek to provide a microscopical foundation for gravity.

\section*{ACKNOWLEDGEMENTS}
F. V. acknowledges support by the Department of Physics of Universidad de los Andes. P. B. is funded by the Beatriz Galindo contract BEAGAL 18/00207 (Spain). A.V is supported by a Melbourne Research Scholarship and the N.D. Goldsworthy Scholarship.

\bibliography{bibliography}

\begin{thebibliography}{50}%
\makeatletter
\providecommand \@ifxundefined [1]{%
 \@ifx{#1\undefined}
}%
\providecommand \@ifnum [1]{%
 \ifnum #1\expandafter \@firstoftwo
 \else \expandafter \@secondoftwo
 \fi
}%
\providecommand \@ifx [1]{%
 \ifx #1\expandafter \@firstoftwo
 \else \expandafter \@secondoftwo
 \fi
}%
\providecommand \natexlab [1]{#1}%
\providecommand \enquote  [1]{``#1''}%
\providecommand \bibnamefont  [1]{#1}%
\providecommand \bibfnamefont [1]{#1}%
\providecommand \citenamefont [1]{#1}%
\providecommand \href@noop [0]{\@secondoftwo}%
\providecommand \href [0]{\begingroup \@sanitize@url \@href}%
\providecommand \@href[1]{\@@startlink{#1}\@@href}%
\providecommand \@@href[1]{\endgroup#1\@@endlink}%
\providecommand \@sanitize@url [0]{\catcode `\\12\catcode `\$12\catcode
  `\&12\catcode `\#12\catcode `\^12\catcode `\_12\catcode `\%12\relax}%
\providecommand \@@startlink[1]{}%
\providecommand \@@endlink[0]{}%
\providecommand \url  [0]{\begingroup\@sanitize@url \@url }%
\providecommand \@url [1]{\endgroup\@href {#1}{\urlprefix }}%
\providecommand \urlprefix  [0]{URL }%
\providecommand \Eprint [0]{\href }%
\providecommand \doibase [0]{http://dx.doi.org/}%
\providecommand \selectlanguage [0]{\@gobble}%
\providecommand \bibinfo  [0]{\@secondoftwo}%
\providecommand \bibfield  [0]{\@secondoftwo}%
\providecommand \translation [1]{[#1]}%
\providecommand \BibitemOpen [0]{}%
\providecommand \bibitemStop [0]{}%
\providecommand \bibitemNoStop [0]{.\EOS\space}%
\providecommand \EOS [0]{\spacefactor3000\relax}%
\providecommand \BibitemShut  [1]{\csname bibitem#1\endcsname}%
\let\auto@bib@innerbib\@empty
\bibitem [{\citenamefont {Penrose}(1972)}]{penrose1972techniques}%
  \BibitemOpen
  \bibfield  {author} {\bibinfo {author} {\bibfnamefont {R.}~\bibnamefont
  {Penrose}},\ }\href@noop {} {\emph {\bibinfo {title} {Techniques of
  differential topology in relativity}}},\ Vol.~\bibinfo {volume} {7}\
  (\bibinfo  {publisher} {Siam},\ \bibinfo {year} {1972})\BibitemShut {NoStop}%
\bibitem [{\citenamefont {Hawking}\ and\ \citenamefont
  {Ellis}(1973)}]{hawking1973large}%
  \BibitemOpen
  \bibfield  {author} {\bibinfo {author} {\bibfnamefont {S.~W.}\ \bibnamefont
  {Hawking}}\ and\ \bibinfo {author} {\bibfnamefont {G.~F.~R.}\ \bibnamefont
  {Ellis}},\ }\href@noop {} {\emph {\bibinfo {title} {The large scale structure
  of space-time}}},\ Vol.~\bibinfo {volume} {1}\ (\bibinfo  {publisher}
  {Cambridge university press},\ \bibinfo {year} {1973})\BibitemShut {NoStop}%
\bibitem [{\citenamefont {Penrose}\ and\ \citenamefont
  {Rindler}(1984{\natexlab{a}})}]{penrose1984spinors}%
  \BibitemOpen
  \bibfield  {author} {\bibinfo {author} {\bibfnamefont {R.}~\bibnamefont
  {Penrose}}\ and\ \bibinfo {author} {\bibfnamefont {W.}~\bibnamefont
  {Rindler}},\ }\href@noop {} {\emph {\bibinfo {title} {Spinors and space-time:
  Volume 1, Two-spinor calculus and relativistic fields}}},\ Vol.~\bibinfo
  {volume} {1}\ (\bibinfo  {publisher} {Cambridge University Press},\ \bibinfo
  {year} {1984})\BibitemShut {NoStop}%
\bibitem [{\citenamefont {Penrose}\ and\ \citenamefont
  {Rindler}(1984{\natexlab{b}})}]{penrose1984spinors2}%
  \BibitemOpen
  \bibfield  {author} {\bibinfo {author} {\bibfnamefont {R.}~\bibnamefont
  {Penrose}}\ and\ \bibinfo {author} {\bibfnamefont {W.}~\bibnamefont
  {Rindler}},\ }\href@noop {} {\emph {\bibinfo {title} {Spinors and space-time:
  Volume 2, Spinor and twistor methods in space-time geometry}}},\
  Vol.~\bibinfo {volume} {2}\ (\bibinfo  {publisher} {Cambridge University
  Press},\ \bibinfo {year} {1984})\BibitemShut {NoStop}%
\bibitem [{\citenamefont {Hawking}(1972)}]{hawking1972black}%
  \BibitemOpen
  \bibfield  {author} {\bibinfo {author} {\bibfnamefont {S.~W.}\ \bibnamefont
  {Hawking}},\ }\href@noop {} {\bibfield  {journal} {\bibinfo  {journal}
  {Communications in Mathematical Physics}\ }\textbf {\bibinfo {volume} {25}},\
  \bibinfo {pages} {152} (\bibinfo {year} {1972})}\BibitemShut {NoStop}%
\bibitem [{\citenamefont {Woolgar}(1999)}]{woolgar1999bounded}%
  \BibitemOpen
  \bibfield  {author} {\bibinfo {author} {\bibfnamefont {E.}~\bibnamefont
  {Woolgar}},\ }\href@noop {} {\bibfield  {journal} {\bibinfo  {journal}
  {Classical and Quantum Gravity}\ }\textbf {\bibinfo {volume} {16}},\ \bibinfo
  {pages} {3005} (\bibinfo {year} {1999})}\BibitemShut {NoStop}%
\bibitem [{\citenamefont {Penrose}(1982)}]{penrose1982quasi}%
  \BibitemOpen
  \bibfield  {author} {\bibinfo {author} {\bibfnamefont {R.}~\bibnamefont
  {Penrose}},\ }\href@noop {} {\bibfield  {journal} {\bibinfo  {journal}
  {Proceedings of the Royal Society of London. A. Mathematical and Physical
  Sciences}\ }\textbf {\bibinfo {volume} {381}},\ \bibinfo {pages} {53}
  (\bibinfo {year} {1982})}\BibitemShut {NoStop}%
\bibitem [{\citenamefont {Hayward}(1994{\natexlab{a}})}]{hayward1994quasi}%
  \BibitemOpen
  \bibfield  {author} {\bibinfo {author} {\bibfnamefont {S.~A.}\ \bibnamefont
  {Hayward}},\ }\href@noop {} {\bibfield  {journal} {\bibinfo  {journal}
  {Classical and Quantum Gravity}\ }\textbf {\bibinfo {volume} {11}},\ \bibinfo
  {pages} {3037} (\bibinfo {year} {1994}{\natexlab{a}})}\BibitemShut {NoStop}%
\bibitem [{\citenamefont {Hayward}(2006)}]{hayward2006gravitational}%
  \BibitemOpen
  \bibfield  {author} {\bibinfo {author} {\bibfnamefont {S.~A.}\ \bibnamefont
  {Hayward}},\ }\href@noop {} {\bibfield  {journal} {\bibinfo  {journal}
  {Classical and Quantum Gravity}\ }\textbf {\bibinfo {volume} {23}},\ \bibinfo
  {pages} {L15} (\bibinfo {year} {2006})}\BibitemShut {NoStop}%
\bibitem [{\citenamefont {Hayward}(1994{\natexlab{b}})}]{hayward1994spin}%
  \BibitemOpen
  \bibfield  {author} {\bibinfo {author} {\bibfnamefont {S.~A.}\ \bibnamefont
  {Hayward}},\ }\href@noop {} {\bibfield  {journal} {\bibinfo  {journal}
  {Classical and Quantum Gravity}\ }\textbf {\bibinfo {volume} {11}},\ \bibinfo
  {pages} {3025} (\bibinfo {year} {1994}{\natexlab{b}})}\BibitemShut {NoStop}%
\bibitem [{\citenamefont {Owen}\ \emph {et~al.}(2011)\citenamefont {Owen},
  \citenamefont {Brink}, \citenamefont {Chen}, \citenamefont {Kaplan},
  \citenamefont {Lovelace}, \citenamefont {Matthews}, \citenamefont {Nichols},
  \citenamefont {Scheel}, \citenamefont {Zhang}, \citenamefont {Zimmerman}
  \emph {et~al.}}]{owen2011frame}%
  \BibitemOpen
  \bibfield  {author} {\bibinfo {author} {\bibfnamefont {R.}~\bibnamefont
  {Owen}}, \bibinfo {author} {\bibfnamefont {J.}~\bibnamefont {Brink}},
  \bibinfo {author} {\bibfnamefont {Y.}~\bibnamefont {Chen}}, \bibinfo {author}
  {\bibfnamefont {J.~D.}\ \bibnamefont {Kaplan}}, \bibinfo {author}
  {\bibfnamefont {G.}~\bibnamefont {Lovelace}}, \bibinfo {author}
  {\bibfnamefont {K.~D.}\ \bibnamefont {Matthews}}, \bibinfo {author}
  {\bibfnamefont {D.~A.}\ \bibnamefont {Nichols}}, \bibinfo {author}
  {\bibfnamefont {M.~A.}\ \bibnamefont {Scheel}}, \bibinfo {author}
  {\bibfnamefont {F.}~\bibnamefont {Zhang}}, \bibinfo {author} {\bibfnamefont
  {A.}~\bibnamefont {Zimmerman}},  \emph {et~al.},\ }\href@noop {} {\bibfield
  {journal} {\bibinfo  {journal} {Physical review letters}\ }\textbf {\bibinfo
  {volume} {106}},\ \bibinfo {pages} {151101} (\bibinfo {year}
  {2011})}\BibitemShut {NoStop}%
\bibitem [{\citenamefont {Nichols}\ \emph {et~al.}(2011)\citenamefont
  {Nichols}, \citenamefont {Owen}, \citenamefont {Zhang}, \citenamefont
  {Zimmerman}, \citenamefont {Brink}, \citenamefont {Chen}, \citenamefont
  {Kaplan}, \citenamefont {Lovelace}, \citenamefont {Matthews}, \citenamefont
  {Scheel} \emph {et~al.}}]{nichols2011visualizing}%
  \BibitemOpen
  \bibfield  {author} {\bibinfo {author} {\bibfnamefont {D.~A.}\ \bibnamefont
  {Nichols}}, \bibinfo {author} {\bibfnamefont {R.}~\bibnamefont {Owen}},
  \bibinfo {author} {\bibfnamefont {F.}~\bibnamefont {Zhang}}, \bibinfo
  {author} {\bibfnamefont {A.}~\bibnamefont {Zimmerman}}, \bibinfo {author}
  {\bibfnamefont {J.}~\bibnamefont {Brink}}, \bibinfo {author} {\bibfnamefont
  {Y.}~\bibnamefont {Chen}}, \bibinfo {author} {\bibfnamefont {J.~D.}\
  \bibnamefont {Kaplan}}, \bibinfo {author} {\bibfnamefont {G.}~\bibnamefont
  {Lovelace}}, \bibinfo {author} {\bibfnamefont {K.~D.}\ \bibnamefont
  {Matthews}}, \bibinfo {author} {\bibfnamefont {M.~A.}\ \bibnamefont
  {Scheel}},  \emph {et~al.},\ }\href@noop {} {\bibfield  {journal} {\bibinfo
  {journal} {Physical Review D}\ }\textbf {\bibinfo {volume} {84}},\ \bibinfo
  {pages} {124014} (\bibinfo {year} {2011})}\BibitemShut {NoStop}%
\bibitem [{\citenamefont {Zhang}\ \emph {et~al.}(2012)\citenamefont {Zhang},
  \citenamefont {Zimmerman}, \citenamefont {Nichols}, \citenamefont {Chen},
  \citenamefont {Lovelace}, \citenamefont {Matthews}, \citenamefont {Owen},\
  and\ \citenamefont {Thorne}}]{zhang2012visualizing}%
  \BibitemOpen
  \bibfield  {author} {\bibinfo {author} {\bibfnamefont {F.}~\bibnamefont
  {Zhang}}, \bibinfo {author} {\bibfnamefont {A.}~\bibnamefont {Zimmerman}},
  \bibinfo {author} {\bibfnamefont {D.~A.}\ \bibnamefont {Nichols}}, \bibinfo
  {author} {\bibfnamefont {Y.}~\bibnamefont {Chen}}, \bibinfo {author}
  {\bibfnamefont {G.}~\bibnamefont {Lovelace}}, \bibinfo {author}
  {\bibfnamefont {K.~D.}\ \bibnamefont {Matthews}}, \bibinfo {author}
  {\bibfnamefont {R.}~\bibnamefont {Owen}}, \ and\ \bibinfo {author}
  {\bibfnamefont {K.~S.}\ \bibnamefont {Thorne}},\ }\href@noop {} {\bibfield
  {journal} {\bibinfo  {journal} {Physical Review D}\ }\textbf {\bibinfo
  {volume} {86}},\ \bibinfo {pages} {084049} (\bibinfo {year}
  {2012})}\BibitemShut {NoStop}%
\bibitem [{\citenamefont {Jacobson}(1995)}]{jacobson1995thermodynamics}%
  \BibitemOpen
  \bibfield  {author} {\bibinfo {author} {\bibfnamefont {T.}~\bibnamefont
  {Jacobson}},\ }\href@noop {} {\bibfield  {journal} {\bibinfo  {journal}
  {Physical Review Letters}\ }\textbf {\bibinfo {volume} {75}},\ \bibinfo
  {pages} {1260} (\bibinfo {year} {1995})}\BibitemShut {NoStop}%
\bibitem [{\citenamefont
  {Padmanabhan}(2010{\natexlab{a}})}]{padmanabhan2010thermodynamical}%
  \BibitemOpen
  \bibfield  {author} {\bibinfo {author} {\bibfnamefont {T.}~\bibnamefont
  {Padmanabhan}},\ }\href@noop {} {\bibfield  {journal} {\bibinfo  {journal}
  {Reports on Progress in Physics}\ }\textbf {\bibinfo {volume} {73}},\
  \bibinfo {pages} {046901} (\bibinfo {year} {2010}{\natexlab{a}})}\BibitemShut
  {NoStop}%
\bibitem [{\citenamefont {Kastor}\ \emph {et~al.}(2009)\citenamefont {Kastor},
  \citenamefont {Ray},\ and\ \citenamefont {Traschen}}]{kastor2009enthalpy}%
  \BibitemOpen
  \bibfield  {author} {\bibinfo {author} {\bibfnamefont {D.}~\bibnamefont
  {Kastor}}, \bibinfo {author} {\bibfnamefont {S.}~\bibnamefont {Ray}}, \ and\
  \bibinfo {author} {\bibfnamefont {J.}~\bibnamefont {Traschen}},\ }\href@noop
  {} {\bibfield  {journal} {\bibinfo  {journal} {Classical and Quantum
  Gravity}\ }\textbf {\bibinfo {volume} {26}},\ \bibinfo {pages} {195011}
  (\bibinfo {year} {2009})}\BibitemShut {NoStop}%
\bibitem [{\citenamefont {Teitelboim}(1985)}]{teitelboim1985cosmological}%
  \BibitemOpen
  \bibfield  {author} {\bibinfo {author} {\bibfnamefont {C.}~\bibnamefont
  {Teitelboim}},\ }\href@noop {} {\bibfield  {journal} {\bibinfo  {journal}
  {Physics Letters B}\ }\textbf {\bibinfo {volume} {158}},\ \bibinfo {pages}
  {293} (\bibinfo {year} {1985})}\BibitemShut {NoStop}%
\bibitem [{\citenamefont {Henneaux}\ and\ \citenamefont
  {Teitelboim}(1989)}]{henneaux1989cosmological}%
  \BibitemOpen
  \bibfield  {author} {\bibinfo {author} {\bibfnamefont {M.}~\bibnamefont
  {Henneaux}}\ and\ \bibinfo {author} {\bibfnamefont {C.}~\bibnamefont
  {Teitelboim}},\ }\href@noop {} {\bibfield  {journal} {\bibinfo  {journal}
  {Physics Letters B}\ }\textbf {\bibinfo {volume} {222}},\ \bibinfo {pages}
  {195} (\bibinfo {year} {1989})}\BibitemShut {NoStop}%
\bibitem [{\citenamefont {Ashtekar}\ and\ \citenamefont
  {Magnon}(1984)}]{ashtekar1984asymptotically}%
  \BibitemOpen
  \bibfield  {author} {\bibinfo {author} {\bibfnamefont {A.}~\bibnamefont
  {Ashtekar}}\ and\ \bibinfo {author} {\bibfnamefont {A.}~\bibnamefont
  {Magnon}},\ }\href@noop {} {\bibfield  {journal} {\bibinfo  {journal}
  {Classical and Quantum Gravity}\ }\textbf {\bibinfo {volume} {1}},\ \bibinfo
  {pages} {L39} (\bibinfo {year} {1984})}\BibitemShut {NoStop}%
\bibitem [{\citenamefont {Kubiz{\v{n}}{\'a}k}\ \emph
  {et~al.}(2017)\citenamefont {Kubiz{\v{n}}{\'a}k}, \citenamefont {Mann},\ and\
  \citenamefont {Teo}}]{kubizvnak2017black}%
  \BibitemOpen
  \bibfield  {author} {\bibinfo {author} {\bibfnamefont {D.}~\bibnamefont
  {Kubiz{\v{n}}{\'a}k}}, \bibinfo {author} {\bibfnamefont {R.~B.}\ \bibnamefont
  {Mann}}, \ and\ \bibinfo {author} {\bibfnamefont {M.}~\bibnamefont {Teo}},\
  }\href@noop {} {\bibfield  {journal} {\bibinfo  {journal} {Classical and
  Quantum Gravity}\ }\textbf {\bibinfo {volume} {34}},\ \bibinfo {pages}
  {063001} (\bibinfo {year} {2017})}\BibitemShut {NoStop}%
\bibitem [{\citenamefont {Dolan}(2011{\natexlab{a}})}]{dolan2011pressure}%
  \BibitemOpen
  \bibfield  {author} {\bibinfo {author} {\bibfnamefont {B.~P.}\ \bibnamefont
  {Dolan}},\ }\href@noop {} {\bibfield  {journal} {\bibinfo  {journal}
  {Classical and Quantum Gravity}\ }\textbf {\bibinfo {volume} {28}},\ \bibinfo
  {pages} {235017} (\bibinfo {year} {2011}{\natexlab{a}})}\BibitemShut
  {NoStop}%
\bibitem [{\citenamefont {Bonilla}\ and\ \citenamefont
  {Senovilla}(1997)}]{bonilla1997some}%
  \BibitemOpen
  \bibfield  {author} {\bibinfo {author} {\bibfnamefont {M.~{\'A}.}\
  \bibnamefont {Bonilla}}\ and\ \bibinfo {author} {\bibfnamefont {J.~M.}\
  \bibnamefont {Senovilla}},\ }\href@noop {} {\bibfield  {journal} {\bibinfo
  {journal} {General Relativity and Gravitation}\ }\textbf {\bibinfo {volume}
  {29}},\ \bibinfo {pages} {91} (\bibinfo {year} {1997})}\BibitemShut {NoStop}%
\bibitem [{\citenamefont {Dolan}(2011{\natexlab{b}})}]{dolan2011cosmological}%
  \BibitemOpen
  \bibfield  {author} {\bibinfo {author} {\bibfnamefont {B.~P.}\ \bibnamefont
  {Dolan}},\ }\href@noop {} {\bibfield  {journal} {\bibinfo  {journal}
  {Classical and Quantum Gravity}\ }\textbf {\bibinfo {volume} {28}},\ \bibinfo
  {pages} {125020} (\bibinfo {year} {2011}{\natexlab{b}})}\BibitemShut
  {NoStop}%
\bibitem [{\citenamefont {Cveti{\v{c}}}\ \emph {et~al.}(2011)\citenamefont
  {Cveti{\v{c}}}, \citenamefont {Gibbons}, \citenamefont {Kubiz{\v{n}}{\'a}k},\
  and\ \citenamefont {Pope}}]{cvetivc2011black}%
  \BibitemOpen
  \bibfield  {author} {\bibinfo {author} {\bibfnamefont {M.}~\bibnamefont
  {Cveti{\v{c}}}}, \bibinfo {author} {\bibfnamefont {G.}~\bibnamefont
  {Gibbons}}, \bibinfo {author} {\bibfnamefont {D.}~\bibnamefont
  {Kubiz{\v{n}}{\'a}k}}, \ and\ \bibinfo {author} {\bibfnamefont
  {C.}~\bibnamefont {Pope}},\ }\href@noop {} {\bibfield  {journal} {\bibinfo
  {journal} {Physical Review D}\ }\textbf {\bibinfo {volume} {84}},\ \bibinfo
  {pages} {024037} (\bibinfo {year} {2011})}\BibitemShut {NoStop}%
\bibitem [{\citenamefont {Dolan}(2012)}]{dolan2012pdv}%
  \BibitemOpen
  \bibfield  {author} {\bibinfo {author} {\bibfnamefont {B.~P.}\ \bibnamefont
  {Dolan}},\ }\href@noop {} {\bibfield  {journal} {\bibinfo  {journal} {Open
  Questions in Cosmology}\ } (\bibinfo {year} {2012})}\BibitemShut {NoStop}%
\bibitem [{\citenamefont {Kubiz{\v{n}}{\'a}k}\ and\ \citenamefont
  {Mann}(2015)}]{kubizvnak2015black}%
  \BibitemOpen
  \bibfield  {author} {\bibinfo {author} {\bibfnamefont {D.}~\bibnamefont
  {Kubiz{\v{n}}{\'a}k}}\ and\ \bibinfo {author} {\bibfnamefont {R.~B.}\
  \bibnamefont {Mann}},\ }\href@noop {} {\bibfield  {journal} {\bibinfo
  {journal} {Canadian Journal of Physics}\ }\textbf {\bibinfo {volume} {93}},\
  \bibinfo {pages} {999} (\bibinfo {year} {2015})}\BibitemShut {NoStop}%
\bibitem [{\citenamefont {Mann}(2016)}]{mann2016chemistry}%
  \BibitemOpen
  \bibfield  {author} {\bibinfo {author} {\bibfnamefont {R.~B.}\ \bibnamefont
  {Mann}},\ }in\ \href@noop {} {\emph {\bibinfo {booktitle} {1st Karl
  Schwarzschild Meeting on Gravitational Physics}}}\ (\bibinfo {organization}
  {Springer},\ \bibinfo {year} {2016})\ pp.\ \bibinfo {pages}
  {197--205}\BibitemShut {NoStop}%
\bibitem [{\citenamefont {Dolan}(2015)}]{dolan2015black}%
  \BibitemOpen
  \bibfield  {author} {\bibinfo {author} {\bibfnamefont {B.~P.}\ \bibnamefont
  {Dolan}},\ }\href@noop {} {\bibfield  {journal} {\bibinfo  {journal} {Modern
  Physics Letters A}\ }\textbf {\bibinfo {volume} {30}},\ \bibinfo {pages}
  {1540002} (\bibinfo {year} {2015})}\BibitemShut {NoStop}%
\bibitem [{\citenamefont {Vargas}\ \emph {et~al.}(2018)\citenamefont {Vargas},
  \citenamefont {Contreras},\ and\ \citenamefont {Bargueno}}]{vargas2018sads}%
  \BibitemOpen
  \bibfield  {author} {\bibinfo {author} {\bibfnamefont {A.}~\bibnamefont
  {Vargas}}, \bibinfo {author} {\bibfnamefont {E.}~\bibnamefont {Contreras}}, \
  and\ \bibinfo {author} {\bibfnamefont {P.}~\bibnamefont {Bargueno}},\
  }\href@noop {} {\bibfield  {journal} {\bibinfo  {journal} {General Relativity
  and Gravitation}\ }\textbf {\bibinfo {volume} {50}},\ \bibinfo {pages} {117}
  (\bibinfo {year} {2018})}\BibitemShut {NoStop}%
\bibitem [{\citenamefont
  {Padmanabhan}(2010{\natexlab{b}})}]{padmanabhan2010equipartition}%
  \BibitemOpen
  \bibfield  {author} {\bibinfo {author} {\bibfnamefont {T.}~\bibnamefont
  {Padmanabhan}},\ }\href@noop {} {\bibfield  {journal} {\bibinfo  {journal}
  {International Journal of Modern Physics D}\ }\textbf {\bibinfo {volume}
  {19}},\ \bibinfo {pages} {2275} (\bibinfo {year}
  {2010}{\natexlab{b}})}\BibitemShut {NoStop}%
\bibitem [{\citenamefont
  {Padmanabhan}(2010{\natexlab{c}})}]{padmanabhan2010equipartition2}%
  \BibitemOpen
  \bibfield  {author} {\bibinfo {author} {\bibfnamefont {T.}~\bibnamefont
  {Padmanabhan}},\ }\href@noop {} {\bibfield  {journal} {\bibinfo  {journal}
  {Modern Physics Letters A}\ }\textbf {\bibinfo {volume} {25}},\ \bibinfo
  {pages} {1129} (\bibinfo {year} {2010}{\natexlab{c}})}\BibitemShut {NoStop}%
\bibitem [{\citenamefont
  {Padmanabhan}(2010{\natexlab{d}})}]{padmanabhan2010surface}%
  \BibitemOpen
  \bibfield  {author} {\bibinfo {author} {\bibfnamefont {T.}~\bibnamefont
  {Padmanabhan}},\ }\href@noop {} {\bibfield  {journal} {\bibinfo  {journal}
  {Physical Review D}\ }\textbf {\bibinfo {volume} {81}},\ \bibinfo {pages}
  {124040} (\bibinfo {year} {2010}{\natexlab{d}})}\BibitemShut {NoStop}%
\bibitem [{\citenamefont {Padmanabhan}(2016)}]{padmanabhan2016momentum}%
  \BibitemOpen
  \bibfield  {author} {\bibinfo {author} {\bibfnamefont {T.}~\bibnamefont
  {Padmanabhan}},\ }\href@noop {} {\bibfield  {journal} {\bibinfo  {journal}
  {General Relativity and Gravitation}\ }\textbf {\bibinfo {volume} {48}},\
  \bibinfo {pages} {4} (\bibinfo {year} {2016})}\BibitemShut {NoStop}%
\bibitem [{\citenamefont {Padmanabhan}(2002)}]{padmanabhan2002classical}%
  \BibitemOpen
  \bibfield  {author} {\bibinfo {author} {\bibfnamefont {T.}~\bibnamefont
  {Padmanabhan}},\ }\href@noop {} {\bibfield  {journal} {\bibinfo  {journal}
  {Classical and Quantum Gravity}\ }\textbf {\bibinfo {volume} {19}},\ \bibinfo
  {pages} {5387} (\bibinfo {year} {2002})}\BibitemShut {NoStop}%
\bibitem [{\citenamefont {Padmanabhan}\ and\ \citenamefont
  {Kothawala}(2013)}]{padmanabhan2013lanczos}%
  \BibitemOpen
  \bibfield  {author} {\bibinfo {author} {\bibfnamefont {T.}~\bibnamefont
  {Padmanabhan}}\ and\ \bibinfo {author} {\bibfnamefont {D.}~\bibnamefont
  {Kothawala}},\ }\href@noop {} {\bibfield  {journal} {\bibinfo  {journal}
  {Physics Reports}\ }\textbf {\bibinfo {volume} {531}},\ \bibinfo {pages}
  {115} (\bibinfo {year} {2013})}\BibitemShut {NoStop}%
\bibitem [{\citenamefont {Chakraborty}\ and\ \citenamefont
  {Padmanabhan}(2015)}]{chakraborty2015thermodynamical}%
  \BibitemOpen
  \bibfield  {author} {\bibinfo {author} {\bibfnamefont {S.}~\bibnamefont
  {Chakraborty}}\ and\ \bibinfo {author} {\bibfnamefont {T.}~\bibnamefont
  {Padmanabhan}},\ }\href@noop {} {\bibfield  {journal} {\bibinfo  {journal}
  {Physical Review D}\ }\textbf {\bibinfo {volume} {92}},\ \bibinfo {pages}
  {104011} (\bibinfo {year} {2015})}\BibitemShut {NoStop}%
\bibitem [{\citenamefont {Hansen}\ \emph {et~al.}(2017)\citenamefont {Hansen},
  \citenamefont {Kubiz{\v{n}}{\'a}k},\ and\ \citenamefont
  {Mann}}]{hansen2017universality}%
  \BibitemOpen
  \bibfield  {author} {\bibinfo {author} {\bibfnamefont {D.}~\bibnamefont
  {Hansen}}, \bibinfo {author} {\bibfnamefont {D.}~\bibnamefont
  {Kubiz{\v{n}}{\'a}k}}, \ and\ \bibinfo {author} {\bibfnamefont {R.~B.}\
  \bibnamefont {Mann}},\ }\href@noop {} {\bibfield  {journal} {\bibinfo
  {journal} {Journal of High Energy Physics}\ }\textbf {\bibinfo {volume}
  {2017}},\ \bibinfo {pages} {47} (\bibinfo {year} {2017})}\BibitemShut
  {NoStop}%
\bibitem [{\citenamefont {Paranjape}\ \emph {et~al.}(2006)\citenamefont
  {Paranjape}, \citenamefont {Sarkar},\ and\ \citenamefont
  {Padmanabhan}}]{paranjape2006thermodynamic}%
  \BibitemOpen
  \bibfield  {author} {\bibinfo {author} {\bibfnamefont {A.}~\bibnamefont
  {Paranjape}}, \bibinfo {author} {\bibfnamefont {S.}~\bibnamefont {Sarkar}}, \
  and\ \bibinfo {author} {\bibfnamefont {T.}~\bibnamefont {Padmanabhan}},\
  }\href@noop {} {\bibfield  {journal} {\bibinfo  {journal} {Physical Review
  D}\ }\textbf {\bibinfo {volume} {74}},\ \bibinfo {pages} {104015} (\bibinfo
  {year} {2006})}\BibitemShut {NoStop}%
\bibitem [{\citenamefont {Maeda}\ and\ \citenamefont
  {Nozawa}(2008)}]{maeda2008generalized}%
  \BibitemOpen
  \bibfield  {author} {\bibinfo {author} {\bibfnamefont {H.}~\bibnamefont
  {Maeda}}\ and\ \bibinfo {author} {\bibfnamefont {M.}~\bibnamefont {Nozawa}},\
  }\href@noop {} {\bibfield  {journal} {\bibinfo  {journal} {Physical Review
  D}\ }\textbf {\bibinfo {volume} {77}},\ \bibinfo {pages} {064031} (\bibinfo
  {year} {2008})}\BibitemShut {NoStop}%
\bibitem [{\citenamefont {Cai}\ \emph {et~al.}(2008)\citenamefont {Cai},
  \citenamefont {Cao}, \citenamefont {Hu},\ and\ \citenamefont
  {Kim}}]{cai2008generalized}%
  \BibitemOpen
  \bibfield  {author} {\bibinfo {author} {\bibfnamefont {R.-G.}\ \bibnamefont
  {Cai}}, \bibinfo {author} {\bibfnamefont {L.-M.}\ \bibnamefont {Cao}},
  \bibinfo {author} {\bibfnamefont {Y.-P.}\ \bibnamefont {Hu}}, \ and\ \bibinfo
  {author} {\bibfnamefont {S.~P.}\ \bibnamefont {Kim}},\ }\href@noop {}
  {\bibfield  {journal} {\bibinfo  {journal} {Physical Review D}\ }\textbf
  {\bibinfo {volume} {78}},\ \bibinfo {pages} {124012} (\bibinfo {year}
  {2008})}\BibitemShut {NoStop}%
\bibitem [{\citenamefont {Hayward}(1998)}]{hayward1998unified}%
  \BibitemOpen
  \bibfield  {author} {\bibinfo {author} {\bibfnamefont {S.~A.}\ \bibnamefont
  {Hayward}},\ }\href@noop {} {\bibfield  {journal} {\bibinfo  {journal}
  {Classical and Quantum Gravity}\ }\textbf {\bibinfo {volume} {15}},\ \bibinfo
  {pages} {3147} (\bibinfo {year} {1998})}\BibitemShut {NoStop}%
\bibitem [{\citenamefont {Bel}(1958)}]{bel1958definition}%
  \BibitemOpen
  \bibfield  {author} {\bibinfo {author} {\bibfnamefont {L.}~\bibnamefont
  {Bel}},\ }\href@noop {} {\bibfield  {journal} {\bibinfo  {journal} {CR Acad.
  Sci. Paris}\ }\textbf {\bibinfo {volume} {246}},\ \bibinfo {pages} {3015}
  (\bibinfo {year} {1958})}\BibitemShut {NoStop}%
\bibitem [{\citenamefont {Bel}(1959)}]{bel1959introduction}%
  \BibitemOpen
  \bibfield  {author} {\bibinfo {author} {\bibfnamefont {L.}~\bibnamefont
  {Bel}},\ }\href@noop {} {\bibfield  {journal} {\bibinfo  {journal} {CR Acad.
  Sci. Paris}\ }\textbf {\bibinfo {volume} {248}},\ \bibinfo {pages} {1297}
  (\bibinfo {year} {1959})}\BibitemShut {NoStop}%
\bibitem [{\citenamefont {Bel}(2000)}]{bel2000radiation}%
  \BibitemOpen
  \bibfield  {author} {\bibinfo {author} {\bibfnamefont {L.}~\bibnamefont
  {Bel}},\ }\href@noop {} {\bibfield  {journal} {\bibinfo  {journal} {General
  Relativity and Gravitation}\ }\textbf {\bibinfo {volume} {32}},\ \bibinfo
  {pages} {2047} (\bibinfo {year} {2000})}\BibitemShut {NoStop}%
\bibitem [{\citenamefont {Schouten}(2013)}]{schouten2013ricci}%
  \BibitemOpen
  \bibfield  {author} {\bibinfo {author} {\bibfnamefont {J.~A.}\ \bibnamefont
  {Schouten}},\ }\href@noop {} {\emph {\bibinfo {title} {Ricci-calculus: an
  introduction to tensor analysis and its geometrical applications}}},\
  Vol.~\bibinfo {volume} {10}\ (\bibinfo  {publisher} {Springer Science \&
  Business Media},\ \bibinfo {year} {2013})\BibitemShut {NoStop}%
\bibitem [{\citenamefont {Acquaviva}\ \emph {et~al.}(2018)\citenamefont
  {Acquaviva}, \citenamefont {Kofro{\v{n}}},\ and\ \citenamefont
  {Scholtz}}]{acquaviva2018gravitational}%
  \BibitemOpen
  \bibfield  {author} {\bibinfo {author} {\bibfnamefont {G.}~\bibnamefont
  {Acquaviva}}, \bibinfo {author} {\bibfnamefont {D.}~\bibnamefont
  {Kofro{\v{n}}}}, \ and\ \bibinfo {author} {\bibfnamefont {M.}~\bibnamefont
  {Scholtz}},\ }\href@noop {} {\bibfield  {journal} {\bibinfo  {journal}
  {Classical and Quantum Gravity}\ }\textbf {\bibinfo {volume} {35}},\ \bibinfo
  {pages} {095001} (\bibinfo {year} {2018})}\BibitemShut {NoStop}%
\bibitem [{\citenamefont {Clifton}\ \emph {et~al.}(2013)\citenamefont
  {Clifton}, \citenamefont {Ellis},\ and\ \citenamefont
  {Tavakol}}]{clifton2013gravitational}%
  \BibitemOpen
  \bibfield  {author} {\bibinfo {author} {\bibfnamefont {T.}~\bibnamefont
  {Clifton}}, \bibinfo {author} {\bibfnamefont {G.~F.}\ \bibnamefont {Ellis}},
  \ and\ \bibinfo {author} {\bibfnamefont {R.}~\bibnamefont {Tavakol}},\
  }\href@noop {} {\bibfield  {journal} {\bibinfo  {journal} {Classical and
  Quantum Gravity}\ }\textbf {\bibinfo {volume} {30}},\ \bibinfo {pages}
  {125009} (\bibinfo {year} {2013})}\BibitemShut {NoStop}%
\bibitem [{\citenamefont {Acquaviva}\ \emph {et~al.}(2015)\citenamefont
  {Acquaviva}, \citenamefont {Ellis}, \citenamefont {Goswami},\ and\
  \citenamefont {Hamid}}]{acquaviva2015constructing}%
  \BibitemOpen
  \bibfield  {author} {\bibinfo {author} {\bibfnamefont {G.}~\bibnamefont
  {Acquaviva}}, \bibinfo {author} {\bibfnamefont {G.~F.}\ \bibnamefont
  {Ellis}}, \bibinfo {author} {\bibfnamefont {R.}~\bibnamefont {Goswami}}, \
  and\ \bibinfo {author} {\bibfnamefont {A.~I.}\ \bibnamefont {Hamid}},\
  }\href@noop {} {\bibfield  {journal} {\bibinfo  {journal} {Physical Review
  D}\ }\textbf {\bibinfo {volume} {91}},\ \bibinfo {pages} {064017} (\bibinfo
  {year} {2015})}\BibitemShut {NoStop}%
\bibitem [{\citenamefont {Hayward}(1993)}]{hayward1993dual}%
  \BibitemOpen
  \bibfield  {author} {\bibinfo {author} {\bibfnamefont {S.~A.}\ \bibnamefont
  {Hayward}},\ }\href@noop {} {\bibfield  {journal} {\bibinfo  {journal}
  {Classical and Quantum Gravity}\ }\textbf {\bibinfo {volume} {10}},\ \bibinfo
  {pages} {779} (\bibinfo {year} {1993})}\BibitemShut {NoStop}%
\bibitem [{\citenamefont {Johnson}(2014)}]{johnson2014holographic}%
  \BibitemOpen
  \bibfield  {author} {\bibinfo {author} {\bibfnamefont {C.~V.}\ \bibnamefont
  {Johnson}},\ }\href@noop {} {\bibfield  {journal} {\bibinfo  {journal}
  {Classical and Quantum Gravity}\ }\textbf {\bibinfo {volume} {31}},\ \bibinfo
  {pages} {205002} (\bibinfo {year} {2014})}\BibitemShut {NoStop}%
\end{thebibliography}%
\end{document}